\documentclass[pra,showpacs,showkeys,amsfonts,amsmath,twocolumn]{revtex4}
\usepackage{bm}
\usepackage{graphicx}
\RequirePackage{mathptm}

\numberwithin{equation}{section}
\newcommand{\si}[1]{\sigma_{#1}}
\newcommand{\sib}{\boldsymbol{\sigma}}
\newcommand{\sa}[2]{\sigma_{#1}^{#2}}
\newcommand{\ip}[2]{\langle \,{#1},\,{#2}\,\rangle}

\newcommand{\ro}{\varrho}
\newcommand{\ras}{\varrho_{\mr{as}}}

\newcommand{\ga}[1]{\gamma_{#1}}

\newcommand{\te}{\vartheta}
\newcommand{\vf}{\varphi}

\newcommand{\I}{\openone}

\newcommand{\fala}[1]{\widetilde{#1}}
\newcommand{\ket}[1]{|{#1}\rangle}
\newcommand{\bra}[1]{\langle {#1} |}

\newcommand{\R}{\mathbb R}

\newcommand{\PP}{\mathbb P}

\newcommand{\cA}{{\mathcal A}}
\newcommand{\cB}{{\mathcal B}}
\newcommand{\Atot}{{\mathcal A}_{\mathrm {tot}}}
\newcommand{\tr}{\mathrm{tr}\,}
\newcommand{\ptr}[1]{\mathrm{tr}_{#1}}
\newcommand{\tl}[1]{\boldsymbol #1}
\newcommand{\mr}[1]{\mathrm{#1}}

\begin{document}
\title{Spontaneous emission can locally create quantum discord out of classical correlations}
\author{ Marek Gw\'{o}\'{z}d\'{z} and Lech Jak{\'o}bczyk
\footnote{ ljak@ift.uni.wroc.pl}
}
 \affiliation{Institute of Theoretical Physics\\ University of
Wroc{\l}aw\\
Plac Maxa Borna 9, 50-204 Wroc{\l}aw, Poland}
\begin{abstract}
We consider local time evolution given by spontaneous emission in
the system of independent two-level atoms. It is shown that all
classically correlated initial states are driven into the states
with transient non-zero quantum discord. Thus local creation of
genuine quantum correlations can be observed in a simple physical
system of non-interacting atoms which are not completely isolated
from the environment.
\end{abstract}
 \pacs{03.67.Mn,03.65.Yz,42.50.-p} \keywords{quantum discord, spontaneous emission}
\maketitle
\section{Introduction}
Quantum discord \cite{Z} is the most promising measure of bipartite
quantum correlations beyond celebrated quantum entanglement.  Some
mixed separable states behave "non-classically" with respect to the
local measurements and exactly such states have non zero quantum
discord. As it was shown recently, almost all quantum states have
non-vanishing discord \cite{F}, so an arbitrary small disturbance
can drive a classically correlated state into a state with genuine
quantum correlations. In contrast to the entanglement, even local
noise can increase or create quantum discord \cite{Str,Hu1} and it
was  shown that a trace-preserving local channel can create quantum
correlations in some classical states if and only if it is not
commutativity preserving channel (see e.g. \cite{Hu2, Guo}).
\par
Although this phenomenon was described in some recent publications
(see e.g. \cite{Ci, Ci1, Ge, Ca}) we reconsider it again in a simple and
natural physical situation. It turns out that to observe such local
creation of quantum correlations it is enough to take two non -
interacting two - level atoms not completely  isolated from the
environment. Time evolution of the system  is given by  a
dissipative process of spontaneous emission. In the Markov
approximation, this time evolution is described by the semi - group
of local completely positive mappings i.e. local quantum channels.
One can check that the channels are not commutativity preserving, so
local creation of quantum correlations can appear. We show that all
classically correlated initial states (so called classical -
classical  and classical - quantum states) are driven by the process
of spontaneous emission into the states with transient non - zero
quantum discord. The created discord is transient because during
time evolution all correlations asymptotically decay. Apart from the
usual spontaneous emission of two atoms, we also consider
"one-sided" spontaneous emission in which only one atom emits
photons and the other is completely decoupled from the environment.
It turns out that in such system, local creation of quantum discord
is more effective.
\par
To quantify non-classical correlations in bipartite states which may
differ from entanglement, we use geometric quantum discord
\cite{DVB} (see also \cite{BB} for other geometric measures of
correlations). Classically correlated quantum states have zero
discord and we take them as the initial states of the studied
evolution. We found it useful to consider also some measure of total
correlation present in the system. In our study we propose to use an
algebraic measure given by so called maximal mutual correlation
$C_{M}$, introduced in the context of the algebraic approach to
quantum non - separability \cite{DGJ}. This quantity is closely
related to correlation distance of the state \cite{Hall} i.e. the
trace norm distance between given state and the tensor product of
its partial traces. It is easy to compute and contains some
information about correlations beyond quantum discord. In
particular, in the case of classically correlated states $C_{M}>0$
and  it encodes the properties of classical probability
distributions defining such states (Sect. II). In our analysis
$C_{M}$ plays only an additional role. It can help to choose
appropriate initial classical state, since  as we show, during the
time evolution caused by local noise,  part of the initial classical
correlations are transformed into genuine quantum correlations and
such creation of quantum discord is effective when the initial value
of $C_{M}$ is large.
\par
The main purpose  of our work is to study time evolution of a
measure of quantum correlations. Geometric measure of quantum
discord is defined in terms of  minimal distance of the given state
from the set of classical states, so  the proper choice of such a
distance is very important. The measure proposed in \cite{DVB} uses
a Hilbert - Schmidt distance in a set of states. In this case the
minimization process is simple and can be realized analytically for
arbitrary two - qubit states. Despite of this feature, geometric
discord based on the Hilbert - Schmidt distance has some unwanted
properties. The most serious problem with such a measure is that it
may increase under local operations on the unmeasured subsystem
\cite{Piani, Tuf}. Fortunately, this defect can be repaired by using
other norm in the set of states. As was shown in Ref. \cite{Paula},
the best choice is to use the trace norm (or Schatten 1-norm) in
place of Hilbert-Schmidt norm to define geometric discord. Such
defined measure will be able to consistently quantify nonclassical
correlations, but it is more difficult to compute. The closed
formula for it is known only in the case of  Bell-diagonal or X-type
two-qubit states \cite{Paula, Cic}. For technical reasons in this
paper we adopt the standard definition of geometric discord based on
the Hilbert-Schmidt norm. We consider the time evolution of the
classical - classical and classical - quantum initial states with
the largest possible value of maximal mutual correlation $C_{M}$ and
we shown that for such states the geometric discord grows from zero
to some maximal value and then asymptotically decays. There is no
danger that such production of quantum correlations is an artificial
effect of sensitivity of the quantum discord to the choice of
distance measure, because the evolution always concerns the measured
subsystem. Moreover, in the case when the unmeasured subsystem also
evolves, the production of correlations is smaller (Sect. III). The
comparison with the local evolution of trace distance discord even
more supports the conclusion that this effect is real. Preliminary
results of the analysis of this problem show that the trace distance
discord is also locally created \cite{J1}.
\section{Geometric measure of quantum discord and maximal mutual
correlation} We start with the introduction of the standard notion
of geometric quantum discord \cite{DVB}. When a $d\otimes d$
bipartite system $AB$ is prepared in a state $\ro$ and we perform
local measurement on the subsystem $A$, almost all  states $\ro$
will be disturbed due to such measurement. The (one-sided) geometric
discord $D_{G}^{A}(\ro)$ can be defined as the minimal disturbance,
measured by the squared Hilbert-Schmidt distance, induced by any
projective measurement $\PP^{A}$ on subsystem $A$ i.e.
\begin{equation}
D_{G}^{A}(\ro)=\frac{d}{d-1}\;\min\limits_{\PP^{A}}\,||\ro-\PP^{A}(\ro)||_{2}^{2},
\label{DG}
\end{equation}
where
\begin{equation}
||m||_{2}=\sqrt{\tr (m\,m^{\dagger})}.
\end{equation}
Here we adopt normalized version of the geometric discord,
introduced in Ref. \cite{Adesso}. There is a subclass of states
which are left unperturbed by at least one measurement. Such
"classical-quantum" states $\ro_{cq}$ satisfy
\begin{equation}
\PP^{A}(\ro_{cq})=\sum\limits_{k}(P_{k}\otimes\I)\,\ro_{cq}\,(P_{k}\otimes
\I)=\ro_{cq}
\end{equation}
for some von Neumann measurement $\{P_{k}\}$. Obviously
$D_{G}^{A}(\ro_{cq})=0$ and
\begin{equation}
\ro_{cq}=\sum\limits_{k}p_{k}P_{k}\otimes \ro^{B}_{k},\label{cq}
\end{equation}
where $P_{k}=\ket{\psi_{k}}\bra{\psi_{k}}$ for some orthonormal
basis $\ket{\psi_{k}}$, $\ro^{B}_{k}$ are quantum states in $B$ and
$\{p_{k}\}$ is the classical probability distribution. The geometric
discord can be also interpreted as the distance between the state
$\ro$ and the closest classical-quantum state. One can also consider
fully classically correlated quantum states (known as
"classical-classical" states), defined as
\begin{equation}
\ro_{cc}=\sum\limits_{j,k}p_{jk}P_{j}^{A}\otimes
P_{k}^{B},\label{cc}
\end{equation}
where $\{P_{j}^{A}\}$ and $\{P_{k}^{B}\}$ are von Neumann
measurements on subsystems $A$ and $B$ respectively, and
$\{p_{jk}\}$ is a two-dimensional probability distribution. The
states (\ref{cc}) satisfy $D_{G}^{A}(\ro_{cc})=0$ but also have zero
discord over two-sided projective measurements \cite{Xu}.
\par
In the case of two qubits studied in the present paper, there is an
explicit expression for $D_{G}^{A}$ \cite{DVB}
\begin{equation}
D_{G}^{A}(\ro)=\frac{1}{2}\,\left(||\tl{x}||^{2}+||T||_{2}^{2}-k_{\mathrm{max}}\right),
\end{equation}
where the components of the vector $\tl{x}\in \R^{3}$ are given by
\begin{equation}
 x_{k}=\tr\, (\ro\,\si{k}\otimes \I).
\end{equation}
The matrix $T$ has elements
\begin{equation}
T_{jk}=\tr\,(\ro\,\si{j}\otimes \si{k})
\end{equation}
and $k_{\mathrm{max}}$ is the largest eigenvalue of the matrix $\tl{x}\,\tl{x}^{T}+T\,T^{T}$.
One also finds a
simple relation between geometric quantum discord and entanglement
measured by normalized negativity
\begin{equation}
N(\ro)=||\ro^{PT}||_{1}-1,\label{neg}
\end{equation}
where $PT$ stands for partial transposition and $||m||_{1}=\tr
\,|m|$. The relation is as follows \cite{Adesso}: for all pure
states
\begin{equation}
\sqrt{D_{G}^{A}(\ket{\psi})}=N(\ket{\psi}),
\end{equation}
whereas for every general two-qubit state $\ro$
\begin{equation}
\sqrt{D_{G}^{A}(\ro)}\geq N(\ro).\label{DvN}
\end{equation}
Notice that the relation (\ref{DvN}) is not universal - there are
conterexamples in higher dimensinal systems \cite{R}.
\par
We also shortly discuss the notion of trace distance geometric
discord $D_{1}^{A}(\ro)$. This quantity is defined as follows
\cite{Paula}
\begin{equation}
D_{1}^{A}(\ro)=\min\limits_{\PP^{A}}\,||\ro-\PP^{A}(\ro)||_{1}.
\end{equation}
In contrast to the geometric discord based on the Hilbert-Schmidt
norm, $D_{1}^{A}$ is nonincreasing under general local operations on
subsystem $B$, but its computation is much more involved \cite{Cic}.
Also the relation between $D_{G}^{A}$ and $D_{1}^{A}$ is not clear.
It can show \cite{Paula} that for Bell-diagonal states
\begin{equation}
D_{1}^{A}(\ro)\geq \sqrt{D_{G}^{A}(\ro)},\label{D1vDG}
\end{equation}
but for general two-qubit states the inequality (\ref{D1vDG}) cannot
be true. In the present paper we study time evolution of
$D_{G}^{A}$, analogous problem for $D_{1}^{A}$ will be considered
elsewhere \cite{J1}.
\par
In a study of time evolution of classical and quantum correlations
in a bipartite systems it is useful to consider some measure of
total correlations present in the system. In this paper we propose
to use so called  maximal mutual correlation $C_{M}(\ro)$, a quantity
motivated by the algebraic approach to quantum entanglement
\cite{DGJ}. In the algebraic formulation, a bipartite quantum system is
defined by the algebra of observables $\Atot$ with distinguished
mutually commuting subalgebras $\cA$ and $\cB$ corresponding to
subsystems. Any state $\ro$ defines a linear functional
$\omega_{\ro}$ on the algebra $\Atot$
\begin{equation}
\omega_{\ro}(a)=\tr (\ro a),\quad a\in \Atot.
\end{equation}
\textit{Maximal mutual correlation} between  the subalgebras $\cA$
and $\cB$ in the state $\ro$ is defined as
\begin{equation}
C_{M}(\ro)=\sup\limits_{a,b}\,|\omega_{\ro}(ab)-\omega_{\ro}(a)\omega_{\ro}(b)|.\label{maxtot}
\end{equation}
In the formula (\ref{maxtot}), the supremum is taken over all
elements $a\in\cA$ and $b\in\cB$ such that $||a||=||b||=1$. The
quantity $C_{M}(\ro)$ indicates how much given state differs from a
product state when we take into account only local observables.
Notice that
\begin{equation}
C_{M}(\ro)\leq ||\omega_{\ro}-\omega_{\ro}^{A}\otimes
\omega_{\ro}^{B}||=||\ro-\ro^{A}\otimes\ro^{B}||_{1}
\end{equation}
where the marginal states $\omega_{\ro}^{A}$ and $\omega_{\ro}^{B}$
are given by partial traces $\ro^{A}$ and $\ro^{B}$ of a density
matrix $\ro$. Thus maximal mutual correlation is always dominated by
the \textit{correlation distance} $C(\ro)$ given by
\begin{equation}
C(\ro)=||\ro-\ro^{A}\otimes \ro^{B}||_{1}
\end{equation}
This quantity was recently introduced by Hall \cite{Hall} in the
context of analysis of quantum mutual information.
\par
For two qubits we may choose
\begin{equation}
a=\tl{a}\cdot \sib\otimes \I,\quad b=\I\otimes \tl{b}\cdot \sib,
\end{equation}
where $\tl{a}$ and $\tl{b}$ are normalized vectors in $\R^{3}$. Then
\begin{equation}
\omega_{\ro}(ab)-\omega_{\ro}(a)\omega_{\ro}(b)=\ip{\tl{a}}{Q\,\tl{b}},
\end{equation}
with the correlation matrix $Q=(q_{ij})$ defined by
\begin{equation}
q_{ij}=\tr (\si{i}\otimes\si{j}\;\ro)-\tr
(\si{i}\otimes\I\;\ro)\,\tr(\I\otimes \si{j}\;\ro).
\end{equation}
Thus
\begin{equation}
C_{M}(\ro)=\sup\limits_{\tl{a},\tl{b}}\,|\ip{\tl{a}}{Q\,\tl{b}}|=||Q||
\end{equation}
and $C_{M}(\ro)$ is given by the matrix norm of the corresponding
correlation matrix i.e. the largest singular value of $Q$. It turns
out that for any pure state $\ket{\psi}$, $C_{M}(\ket{\psi})$ gives
the same information as any other measure of quantum correlations.
In particular \cite{DGJ}
\begin{equation}
C_{M}(\ket{\psi})=\sqrt{D_{G}^{A}(\ket{\psi})}=N(\ket{\psi}).
\end{equation}
For mixed states we expect that
\begin{equation}
C_{M}(\ro)\geq \sqrt{D_{G}^{A}(\ro)}\geq N(\ro).\label{nierow}
\end{equation}
The second inequality was shown in Ref. \cite{Adesso} whereas the
first one can be proved analytically for Bell-diagonal states and
numerical simulations suggest that it is valid for all mixed
two-qubit states \cite{J}. The first inequality also indicates that
$C_{M}$ contains some information about  correlations beyond quantum
discord i.e. classical correlations present in the quantum states.
For fully classically correlated states $\ro_{cc}$  given by
(\ref{cc}) one can show that \cite{J}
\begin{equation}
\begin{split}
C_{M}(\ro_{cc})=&\big|(p_{11}-p_{22})^{2}-(p_{12}-p_{21})^{2}\\
&+p_{12}+p_{21}-p_{11}-p_{22}\big|
\end{split}\label{klasik}
\end{equation}
The right hand side of the equation (\ref{klasik}) can be
interpreted as the modulus of covariance of two discrete random
variables $X$ and $Y$ with values in the set $\{-1,+1\}$ and with
join probability distribution defined by probabilities $p_{jk}$ as
follows
\begin{equation}
\begin{split}
&\mathrm{Prob}\{X=+1,Y=+1\}=p_{11},\\
&\mathrm{Prob}\{X=+1,Y=-1\}=p_{12},\\
&\mathrm{Prob}\{X=-1,Y=+1\}=p_{21},\\
&\mathrm{Prob}\{X=-1,Y=-1\}=p_{22}.
\end{split}
\end{equation}
So indeed $\ro_{cc}$ is only classically correlated and
\begin{equation}
C_{M}(\ro_{cc})=|\mathrm{Cov}(X,Y)|.
\end{equation}
On the other hand, for classical-quantum states $\ro_{cq}$ given by
(\ref{cq}), we obtain
\begin{equation}
C_{M}(\ro_{cq})=2\,p_{1}p_{2}\,||\tl{a}_{1}-\tl{a}_{2}||,\label{klqu}
\end{equation}
where $\tl{a}_{1,2}$ are the Bloch vectors corresponding to the
states $\ro_{1,2}^{B}$ i.e.
\begin{equation}
\ro_{1,2}^{B}=\frac{1}{2}\left(\I+\tl{a}_{1,2}\cdot
\sib\right).\label{roB}
\end{equation}
The factor $2\,p_{1}p_{2}$ in equation (\ref{klqu}) can be connected
with statistical properties of discrete random variable $X$ with
values in the set $\{-1,+1\}$ and probability distribution
\begin{equation}
\mathrm{Prob}\{X=+1\}=p_{1},\quad \mathrm{Prob}\{X=-1\}=p_{2}.
\end{equation}
One checks that
\begin{equation}
2\,p_{1}p_{2}=\frac{1}{2}\,\left(\mathrm{Var} X\right)^{2},
\end{equation}
so
\begin{equation}
C_{M}(\ro_{cq})=\frac{1}{2}\,\left(\mathrm{Var}
X\right)^{2}\,||\tl{a}_{1}-\tl{a}_{2}||.\label{cmrocq}
\end{equation}
In that case maximal mutual correlation depends not only on the
classical probability distribution, but also on the properties of
quantum states of the subsystem $B$.
\par
In the next section we will study time evolution of initial states
$\ro_{cq}$ and $\ro_{cc}$ in the system of two-level atoms. We
consider only spontaneous emission of independent atoms and show
that this simple and natural physical process leads to the transient
creation of  quantum correlations out of classical
correlations.
\section{Spontaneous emission in a system of two-level atoms and local generation of quantum discord}
Consider a system of two independent two - level atoms (atom $A$ and
atom $B$) interacting with environment at zero temperature. When we
take into account only the dissipative process of spontaneous
emission, the dynamics of the system is given by the master equation
\cite{Agarwal}
\begin{equation}
\frac{d\ro}{dt}=L_{AB}\ro,\quad L_{AB}=L_{A}+L_{B},\label{me}
\end{equation}
where for $k=A,\, B$
\begin{equation}
L_{k}=\frac{\ga{0}}{2}\,\left(2\,\sa{-}{k}\ro\sa{+}{k}-\sa{+}{k}\sa{-}{k}\ro-\ro\sa{+}{k}\sa{-}{k}\right).
\end{equation}
In the above equation $\sa{\pm}{A}=\si{\pm}\otimes \I,\,
\sa{\pm}{B}=\I\otimes \si{\pm}$ and $\ga{0}$ is the single atom
spontaneous emission rate. Time evolution of the initial density
matrix $\ro$ of the system is given by the semi-group
$\{T_{t}^{AB}\}_{t\geq 0}$ generated by $L_{AB}$ and such evolution
has a well known properties. Let us remind here the matrix elements
of the state $\ro(t)$ with respect to the basis $\ket{e}_{A}\otimes
\ket{e}_{B},\ket{e}_{A}\otimes \ket{g}_{B}, \ket{g}_{A}\otimes
\ket{e}_{B}, \ket{g}_{A}\otimes \ket{g}_{B}$ (where
$\ket{g}_{k},\ket{e}_{k},\, k=A,B$ are the ground states and excited
states of atoms $A$ and $B$)
\begin{equation}
\begin{split}
&\ro_{11}(t)=e^{-2\ga{0}t}\,\ro_{11},\\
&\ro_{1k}(t)=e^{-\frac{3\ga{0}}{2}t}\,\ro_{1k},\; k=2,3,\\
&\ro_{14}(t)=e^{-\ga{0}t}\,\ro_{14},\\
&\ro_{kk}(t)=\left(e^{-\ga{0}t}-e^{-2\ga{0}t}\right)\,\ro_{11}+e^{-\ga{0}t}\,\ro_{kk},\;k=2,3,\\
&\ro_{23}(t)=\left(e^{-\frac{\ga{0}}{2}t}-e^{-\frac{3\ga{0}}{2}t}\right)\,\ro_{13}+
e^{-\frac{\ga{0}}{2}t}\,\ro_{24},\\
&\ro_{34}(t)=\left(e^{-\frac{\ga{0}}{2}t}-e^{-\frac{3\ga{0}}{2}t}\right)\,\ro_{12}+
e^{-\frac{\ga{0}}{2}t}\,\ro_{34}
\end{split}
\end{equation}
and
\begin{equation}
\ro_{44}(t)=1-\ro_{11}(t)-\ro_{22}(t)-\ro_{33}(t).
\end{equation}
The remaining matrix elements can be obtained by the hermiticity
condition. The evolution given by the semi-group
$\{T_{t}^{AB}\}_{t\geq 0}$ is ergodic: for any initial state, the
asymptotic state is equal to the ground state of two atoms.
\par
 To investigate the process of local generation of quantum discord,
it is instructive to consider "one-sided" spontaneous emission i.e.
time evolution $\{T_{t}^{A}\}_{t\geq 0}$ generated only by $L_{A}$.
Here the atom $A$ spontaneously emits photons, whereas the atom $B$
is isolated from the environment. Such dynamics has  peculiar
properties, so we discuss it in some details. In that case the state
$\ro(t)$ at time $t$ has the following matrix elements
\begin{equation}
\begin{split}
&\ro_{11}(t)=e^{-\ga{0}\,t}\,\ro_{11},\quad
\ro_{12}(t)=e^{-\ga{0}\,t}\,\ro_{12},\\
&\ro_{22}(t)=e^{-\ga{0}\,t}\,\ro_{22},\quad
\ro_{13}(t)=e^{-\frac{\ga{0}}{2}\,t}\,\ro_{13},\\
&\ro_{14}(t)=e^{-\frac{\ga{0}}{2}\,t}\,\ro_{14},\quad
\ro_{23}(t)=e^{-\frac{\ga{0}}{2}\,t}\,\ro_{23},\\
&\ro_{24}(t)=e^{-\frac{\ga{0}}{2}\,t}\,\ro_{24}
\end{split}\label{solution1}
\end{equation}
and
\begin{equation}
\begin{split}
&\ro_{33}(t)=(1-e^{-\ga{0}\,t})\,\ro_{11}+\ro_{33},\\
&\ro_{34}(t)=(1-e^{\ga{0}\,t})\,\ro_{12}+\ro_{34},\\
&\ro_{44}(t)=(1-e^{-\ga{0}\,t})\,\ro_{22}+\ro_{44}.
\end{split}\label{solution2}
\end{equation}
Again the remaining matrix elements can be obtained by the
hermiticity condition.
\par
 From the equations (\ref{solution1}) and
(\ref{solution2}) it follows that in contrast to the usual process
of spontaneous emission, now there are non-trivial asymptotic states
of the system, given by
\begin{equation}
\ras=P_{g}\otimes \ptr{A}\,\ro,
\end{equation}
where $P_{g}$ is the projection on the ground state of the atom $A$
and $\ro$ is the initial density matrix.
\par
Notice that in both cases, asymptotic state is a product state, so
any measure of  quantum correlations should asymptotically vanish
during such time evolutions. But the transient quantum correlation
of some states still can exist. To show this in an explicit way, consider the initial states
$\ro_{cq}$  and $\ro_{cc}$. In the case of two qubits, these states are defined in terms
of orthogonal projectors
\begin{equation}
\begin{split}
&P_{1}=\begin{pmatrix} \cos^{2}\te&\frac{1}{2}e^{-i\vf}\,\sin 2\te\\[2mm]
\frac{1}{2}e^{i\vf}\,\sin 2\te&\sin^{2}\te\end{pmatrix},\\[2mm]
&P_{2}=\begin{pmatrix} \sin^{2}\te&-\frac{1}{2}e^{-i\vf}\,\sin 2\te\\[2mm]
-\frac{1}{2}e^{i\vf}\,\sin 2\te&\cos^{2}\te\end{pmatrix}.
\end{split}\label{P1P2}
\end{equation}
In particular
\begin{equation}
\ro_{cq}=p_{1}\,P_{1}\otimes \ro_{1}^{B}+p_{2}\,P_{2}\otimes
\ro_{2}^{B},\label{rocq2}
\end{equation}
where the states $\ro_{1,2}^{B}$ are given by (\ref{roB}). Similarly, up to the local unitary
equivalence, the states $\ro_{cc}$ can be written as
\begin{equation}
\ro_{cc}=\sum\limits_{j,k=1}^{2}p_{jk}\,P_{j}\otimes
P_{k}.\label{rocc2}
\end{equation}
\par
Since local dissipative evolution can only degrade maximal mutual
correlations and  the relation (\ref{nierow}) is valid for all
times, the initial state should have a large value of $C_{M}$.  In
the case of classically correlated states (\ref{rocc2}), the largest
value of $C_{M}(\ro_{cc})$ is obtained  when for example
\begin{equation*}
p_{11}=p_{22}=0\quad\text{and}\quad p_{12}=p_{21}=\frac{1}{2}.
\end{equation*}
Detailed analysis of time evolution of such initial states shows
that the maximal creation of quantum correlations happens when in
the parametrization of orthogonal projections (\ref{P1P2})  we take
$\te=\pi/4$. So we first consider time evolution of the  initial
state
\begin{equation}
\ro_{0}=\frac{1}{2}\,\ket{+}\bra{+}\otimes
\ket{-}\bra{-}+\frac{1}{2}\,\ket{-}\bra{-}\otimes\ket{+}\bra{+},\label{rocc1}
\end{equation}
where
\begin{equation}
\ket{\pm}=\frac{1}{\sqrt{2}}\,\left(\ket{e}\pm\,\ket{g}\right).
\end{equation}
We study evolution of (\ref{rocc1}) given by the semi-group
$\{T_{t}^{AB}\}_{t\geq 0}$ as well as one-sided evolution defined by
$\{T_{t}^{A}\}_{t\geq 0}$ i.e.
\begin{equation}
\ro_{0}(t)=T_{t}^{AB}\ro_{0},\quad
\fala{\ro}_{0}(t)=T_{t}^{A}\ro_{0}.
\end{equation}
In both cases, the maximal mutual correlation decays
\begin{equation}
C_{M}(\ro_{0}(t))=e^{-\ga{0}\,t},\quad
C_{M}(\fala{\ro}_{0}(t))=e^{-\ga{0}\,t/2},
\end{equation}
but this decay is slower for one-sided spontaneous emission. On the
other hand
\begin{equation}
D_{G}^{A}(\ro_{0}(t))=\min\,\left(D_{1}(t),\,D_{2}(t)\right)
\end{equation}
with
\begin{equation}
\begin{split}
&D_{1}(t)=\frac{1}{2}\,e^{-2\ga{0}\,t},\\
&D_{2}(t)=\frac{1}{2}\,\left[\,(e^{-\ga{0}\,t}-1)^{2}+e^{-4\ga{0}\,t}\,(e^{\ga{0}\,t}-1)^{4}\right]
\end{split}
\end{equation}
and
\begin{equation}
D_{G}^{A}(\fala{\ro}_{0}(t))=\min\,\left(\fala{D}_{1}(t),\,
\fala{D}_{2}(t)\,\right)
\end{equation}
where
\begin{equation}
\fala{D}_{1}(t)=\frac{1}{2}\,e^{-\ga{0}\,t},\quad
\fala{D}_{2}(t)=\frac{1}{2}\,(e^{-\ga{0}\,t}-1)^{2}.
\end{equation}
As we see, one-sided geometric discord behaves differently then $C_{M}$ during
time evolution. In both cases $D_{G}^{A}$ grows from zero to some
maximal value and then it asymptotically goes to zero. Moreover, for
all $t$
\begin{equation*}
C_{M}(\ro_{0}(t))\geq \sqrt{D_{G}^{A}(\ro_{0}(t))}
\end{equation*}
and similarly for one-sided spontaneous emission. Notice also that
one-sided spontaneous emission produces more quantum correlations
then the usual process of local spontaneous emission of two atoms
(FIG.1). It can be connected with the fact that the time evolution operator is
applied only to the part of the system which is measured. On the other hand,
one - sided evolution operating only on subsystem $B$ cannot produce non - zero
(one - sided) quantum discord $D_{G}^{A}$, since initial classical - classical
 or classical - quantum states become classical - quantum states under the evolution.
\begin{figure}[h]
\centering {\includegraphics[height=54mm]{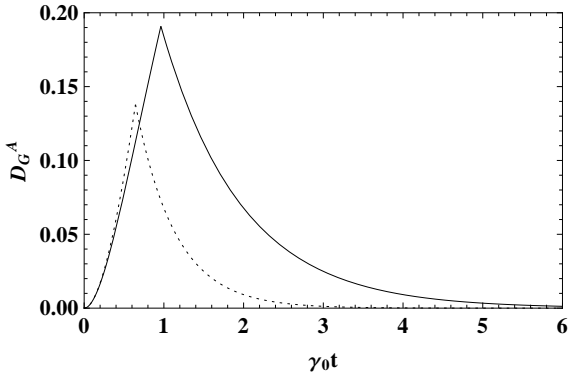}}\caption{Time
evolution of the geometric discord of the initial state $\ro_{0}$
due to one-sided spontaneous emission (solid line) and usual
spontaneous emission of two atoms (dotted line).}
\end{figure}
For other initial states of this type, the analysis of of time
evolution of geometric discord is much more involved. Here we
consider only the initial states $\ro_{1}$ and $\ro_{2}$ defined by
 $\te=\pi/6$ and $\te=\pi/8$, respectively. Figure 2 shows
evolution of $D_{G}^{A}$ for such initial states, compared to the
analogous evolution for the initial state $\ro_{0}$.
\begin{figure}[t]
\centering{\includegraphics[height=54mm]{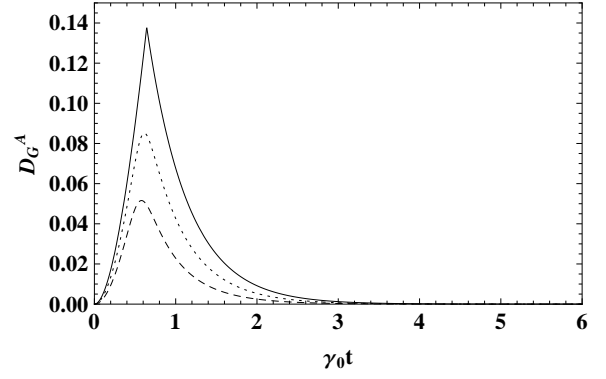}}\caption{Evolution
of $D_{G}^{A}$ for initial states $\ro_{0}$ (solid line), $\ro_{1}$
(dotted line) and $\ro_{2}$ (dashed line).}
\end{figure}
We see that  for such initial states, the creation of quantum
discord is less effective.
\par
Consider now $\ro_{cq}$ as the initial state. It follows from the
formula (\ref{cmrocq}) that for such state the largest  value of
$C_{M}$ is obtained when $p_{1}=p_{2}=1/2$ and the Bloch vectors
$\tl{a}_{1}$ and $\tl{a}_{2}$ are opposite and normalized  i.e. the
states $\ro_{1}^{B}$ and $\ro_{2}^{B}$ are orthogonal projectors.
Such $\ro_{cq}$ is the specific example of classical - classical
state. For the typical classical - quantum state, $C_{M}$ is always
less then $1$. Direct calculations also show that similarly as in
the classical - classical case, the creation of quantum correlations
is maximal when the projectors in the "classical" part of $\ro_{cq}$
are chosen to be
\begin{equation*}
P_{1}=\ket{+}\bra{+},\quad P_{2}=\ket{-}\bra{-}
\end{equation*}
and the initial state is the following
\begin{equation}
\ro=\frac{1}{2}\,\ket{+}\bra{+}\otimes\,
\ro_{1}^{B}+\frac{1}{2}\,\ket{-}\bra{-}\otimes \,\ro_{2}^{B}.
\end{equation}
The creation of quantum discord still depends on the quantum part of
the state $\ro$. To study this dependence, consider the states of
the subsystem $B$ defined by normalized Bloch vectors
\begin{equation*}
\tl{a}_{1}=(\cos\alpha_{0},\sin\alpha_{0},0),\quad \tl{a}_{2}=(\cos
(\alpha_{0}+\alpha),\sin (\alpha_{0}+\alpha),0).
\end{equation*}
Notice that
\begin{equation*}
||\tl{a}_{1}-\tl{a}_{2}||^{2}=2\,(1-\cos\alpha),
\end{equation*}
where the parameter $\alpha_{0}$ is fixed. For such initial states and dynamics given by
spontaneous emission of two atoms, one obtains
\begin{equation}
D_{G}^{A}(\ro(t))=\min\,\left(D_{1}^{q}(t),\, D_{2}^{q}(t)\right)
\end{equation}
with
\begin{equation}
\begin{split}
D_{1}^{q}(t)=&\frac{1}{4}\,(1-\cos\alpha)\,e^{-2\ga{0}\,t},\\
D_{2}^{q}(t)=&1+\frac{1}{2}e^{-4\ga{0}\,t}-\frac{7}{4}e^{-3\ga{0}\,t}+3\,e^{-2\ga{0}\,t}-\frac{11}{4}e^{-\ga{0}\,t}\\
&+\left(\frac{1}{4}e^{-3\ga{0}\,t}-\frac{1}{2}e^{-2\ga{0}\,t}+\frac{1}{4}e^{-\ga{0}\,t}\right)\,\cos\alpha
.
\end{split}
\end{equation}
Again $D_{G}^{A}(\ro(t))$ grows from zero to the maximal value
$D_{\mathrm{max}}$ and then decays to zero. The maximum of the
created discord depends only on the angle $\alpha$
 between the Bloch vectors $\tl{a}_{1}$ and $\tl{a}_{2}$ (or a distance between them)
 i.e. in contrast to the classical part of $\ro$ it does not depend on detailed
structure of the states on the subsystem $B$ but only on the Hilbert
- Schmidt distance between them. $D_{\mathrm{max}}$ as a function of
$\alpha$ has its largest value for $\alpha=\pi$ i.e. for classical -
classical initial state (FIG.3). For $\alpha=0$ or $2\pi$, the produced
discord is zero, since the corresponding initial states are product states
and local time evolution cannot create any quantum correlations in the system \cite{Hu1}.
From this plot we see also that all
classical - quantum states of this type with non - zero distance
between $\ro_{1}^{B}$ and $\ro_{2}^{B}$ (i.e. with non - zero value
of $C_{M}$) evolve into states with non - trivial quantum
correlations. General class of classical - quantum states is much
more difficult to study, but the numerical calculations suggests the
same behavior of geometric discord.
\begin{figure}[t]
\centering{\includegraphics[height=54mm]{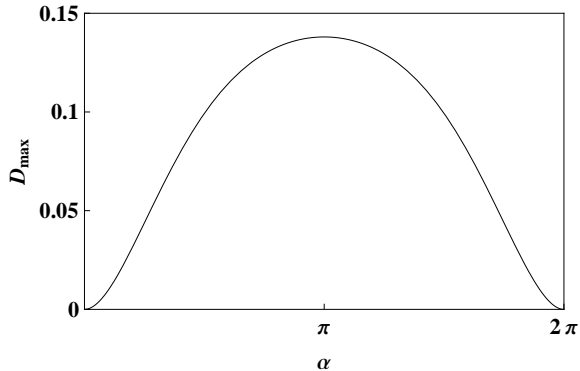}}\caption{Maximal
value of created discord versus the angle between Bloch vectors.}
\end{figure}
\section{Conclusions}
We have investigated the possibility  of local creation of quantum discord in the system
of independent two - level atoms coupled to the vacuum. Individual spontaneous emission of
the atoms leads to the local  evolution destroying quantum entanglement. At the same time,
this kind of evolution can create quantum discord of separable states. We have shown that
all classically correlated initial states are driven by the process of spontaneous emission
into the states which are still separable but have non - zero quantum discord. The largest value
of the created discord is obtained for the classical - classical states (\ref{rocc2}) for which
$C_{M}(\ro_{cc})=1$ and the corresponding orthogonal projectors (\ref{P1P2}) are defined by
$\te=\pi/4$. For the classical - quantum states (\ref{rocq2}) the creation of quantum discord
is always less effective.

\end{document}